\documentclass[twocolumn,english,aps,prl,showpacs]{revtex4}
\usepackage[T1]{fontenc}
\usepackage[latin9]{inputenc}
\usepackage{graphicx}

\makeatletter

\makeatletter

\usepackage{bm}

\makeatletter
\makeatother

\makeatother

\makeatother

\usepackage{babel}

\begin{document}

\title{Unconventional pairing in the iron arsenide superconductors}

\author{Rafael M. Fernandes, Daniel K. Pratt$,$ Wei Tian, Jerel Zarestky,
Andreas Kreyssig, Shibabrata Nandi, Min Gyu Kim, Alex Thaler, Ni Ni,
Paul C. Canfield, Robert J. McQueeney, J\"org Schmalian, and Alan I.
Goldman }

\affiliation{Ames Laboratory, U.S. DOE, and Department of Physics and Astronomy,
Iowa State University, Ames, IA 50011, USA }

\date{\today}

\begin{abstract}
We use magnetic long range order as a tool to probe the Cooper pair
wave function in the iron arsenide superconductors. We show theoretically
that antiferromagnetism and superconductivity can coexist in these
materials only if Cooper pairs form an unconventional, sign-changing
state. The observation of coexistence in Ba(Fe$_{1-x}$Co$_{x}$)$_{2}$As$_{2}$
then demonstrates unconventional pairing in this material. The detailed
agreement between theory and neutron diffraction experiments, in particular
for the unusual behavior of the magnetic order below $T_{c}$, demonstrates the robustness
of our conclusions. Our findings strongly suggest that superconductivity
is unconventional in all members of the iron arsenide family. 
\end{abstract}


\maketitle
The determination of the mechanism of superconductivity in the recently
discovered\cite{Kamihara08,Rotter08} iron arsenide compounds remains
the prime issue in this field, requiring knowledge of the symmetry
and of the internal structure of the Cooper pair wave function. A
promising candidate for the pairing symmetry is the $s^{+-}$-state,
proposed by electronic theories for superconductivity\cite{Mazin08,Kuroki08,Chubukov08,Mazin09},
where the Cooper pair wave function changes sign between different
sheets of the Fermi surface. In distinction to the case of $d$-wave
pairing\cite{Dale95,Tsuei00} in the much studied cuprates, no additional
symmetry is broken for $s^{+-}$-pairing \cite{Tesanovic08}, making
proposals for the determination of the wave function through interference
experiments\cite{Parker09,Wu09,Linder09} more complex and less conclusive.
Important clues about superconductivity in strongly correlated electron
systems can, however, be deduced by investigating their phase diagrams
and the competition between different phases \cite{Sachdev02,Zhang97}. 

The iron arsenide superconductors manifest a rich phase diagram where
antiferromagnetic (AFM), tetragonal (Tet), orthorhombic (Ort) and
superconducting (SC) order are found in close proximity\cite{Luetkens,Park09,Goko09,Ni09,Chu09,Ning09,Lester09}.
For some compounds, the transition between AFM and SC is of first
order\cite{Luetkens,Park09,Goko09} with regions of inhomogeneous
phase coexistence. However, in Ba(Fe$_{1-x}$Co$_{x}$)$_{2}$As$_{2}$,
experiments\cite{Ni09,Chu09,Ning09,Lester09,Laplace09,Julien09,Bernhard09,Pratt09,Christianson09}
have established homogeneous coexistence of SC and AFM for intermediate
$x$-values. As the system enters the SC state, the ordered magnetic
moment smoothly decreases with decreasing temperature\cite{Pratt09,Christianson09},
a behavior in sharp contrast to what is known for many conventional
superconductors \cite{Baltensperger63}, where AFM, associated with
localized spins, can easily coexist with SC. This provides strong
evidence for the fact that superconductivity and magnetic long-range
order compete for the same electrons.

In this Rapid Communication we demonstrate that the Cooper pair wave function in
the iron arsenides is revealed via the coexistence and competition
between superconductivity and magnetic order. We find that AFM and
conventional phonon-mediated SC can not coexist, while unconventional
$s^{+-}$-pairing is located near the borderline between phase coexistence
and mutual exclusion. Therefore, the two phases can coexist only if
Cooper pairing is unconventional with a sign-changing pairing wave
function, whereas the absence of coexistence in other pnictide superconductors
can not be used as evidence for conventional pairing. Our neutron
scattering measurements confirm all aspects of our theory, including
the novel re-entrance of the non-magnetically ordered phase. These
findings strongly suggest that superconductivity is unconventional
in all members of the iron arsenide family, given their similar electronic
structure and transition temperatures.

\emph{Microscopic model:} we use a few basic ingredients to describe
the main features of the iron arsenides: the electronic structure
is characterized by two sets of Fermi surface sheets, a hole pocket
around the center of the Brillouin zone and an electron pocket shifted
by the ordering vector $\mathbf{Q}$ with Hamiltonian:\begin{equation}
H=\sum_{\mathbf{p,}\sigma,l}\varepsilon_{\mathbf{p,}l}\psi_{\mathbf{p}\sigma l}^{\dagger}\psi_{\mathbf{p}\sigma l}+H_{\mathrm{int}}\:.\end{equation}
 We use a circular hole Fermi surface, with $\varepsilon_{\mathbf{p,}1}=\varepsilon_{1,0}-p^{2}/(2m)-\mu$,
and an elliptical electron Fermi surface, with $\varepsilon_{\mathbf{p+Q,}2}=-\varepsilon_{2,0}+p_{x}^{2}/(2m_{x})+p_{y}^{2}/(2m_{y})-\mu$.
For the electron-electron interaction, $H_{\mathrm{int}}$, we include
a magnetic electronic interaction $I$, i.e. $I\sum_{\mathbf{p,p}^{\prime},\mathbf{q}}\psi_{\mathbf{p}s1}^{\dagger}\boldsymbol{\sigma}_{ss^{\prime}}\psi_{\mathbf{p+q}s^{\prime}2}\cdot \psi_{\mathbf{p}^{\prime}s2}^{\dagger}\boldsymbol{\sigma}_{ss^{\prime}}\psi_{\mathbf{p}^{\prime}-\mathbf{q}s^{\prime}1}$
and a pairing interaction $V_{ll^{\prime}}$, i.e. $\sum_{\mathbf{p,p}^{\prime},\mathbf{q,}ll^{\prime}}V_{ll^{\prime}}\psi_{\mathbf{p+q}\mathbf{\uparrow}l}^{\dagger}\psi_{-\mathbf{p\downarrow}l}^{\dagger}\psi_{-\mathbf{p}^{\prime}-\mathbf{q\downarrow}l^{\prime}}\psi_{\mathbf{p}^{\prime}\uparrow l^{\prime}}$.
Although our key results are valid for arbitrary pairing matrix $V_{ll^{\prime}}$,
hereafter we will focus on the case of a predominant interband pairing
$V_{ll^{\prime}}=V\left(1-\delta_{ll^{\prime}}\right)$. Depending
on the choice for the sign of $V$, we consider the $s^{+-}$ state,
as arising from an electronic pairing mechanism with $V>0$\cite{Mazin08,Kuroki08,Chubukov08,Mazin09},
or the $s^{++}$ state that would result from electron-phonon interaction
($V<0$). In the latter case, the Cooper pair wave function has the
same sign in all Fermi sheets. We analyze the resulting model within
a weak coupling mean field theory\cite{Machida81,Vorontsov09} and
obtain the free energy density of a superconductor with antiferromagnetic
long range order: \begin{eqnarray}
F\left(\mathbf{M},\Psi_{\alpha}\right) & = & I\mathbf{M}^{2}-\frac{V}{2}\left(\Psi_{1}^{\ast}\Psi_{2}+\Psi_{2}^{\ast}\Psi_{1}\right)\nonumber \\
 &  & -\frac{T}{N}\sum_{\mathbf{p,}a=\pm}\log\left(4\cosh\left(\frac{E_{\mathbf{p}a}}{2k_{B}T}\right)\right)\:.\label{F}\end{eqnarray}
 The SC order parameters $\Psi_{1}$ and $\Psi_{2}$ of the two bands
and the staggered moment $M$ are obtained by minimizing $F\left(\mathbf{M},\Psi_{\alpha}\right)$,
where $N$ is the system size and $E_{\mathbf{p}a}$ are the positive
eigenvalues of a state with AFM and SC order: \begin{equation}
E_{\mathbf{p}\pm}^{2}=\frac{1}{2}\left(\Gamma_{\mathbf{p}}\pm\sqrt{\Gamma_{\mathbf{p}}^{2}+\Omega_{\mathbf{p}}+\delta_{\mathbf{p}}}\right)\:,\end{equation}
 with $\Gamma_{\mathbf{p}}=2\Delta_{\mathrm{AFM}}^{2}+\Delta_{1}^{2}+\Delta_{2}^{2}+\varepsilon_{\mathbf{p,}1}^{2}+\varepsilon_{\mathbf{p+Q,}2}^{2}$
and $\Omega_{\mathbf{p}}=-4\left(\varepsilon_{\mathbf{p,}1}^{2}+\Delta_{1}^{2}\right)\left(\varepsilon_{\mathbf{p+Q,}2}^{2}+\Delta_{2}^{2}\right)$
as well as $\delta_{\mathbf{p}}=8\Delta_{\mathrm{AFM}}^{2}\left(\Delta_{1}\Delta_{2}+\varepsilon_{\mathbf{p,}1}\varepsilon_{\mathbf{p+Q,}2}-\Delta_{\mathrm{AFM}}^{2}/2\right)$.
Here $\Delta_{l}=V\Psi_{l}$ and $\Delta_{\mathrm{AFM}}=IM$ refer
to the SC and AFM single particle gaps.

\begin{figure}
\begin{centering}
\includegraphics[width=1\columnwidth]{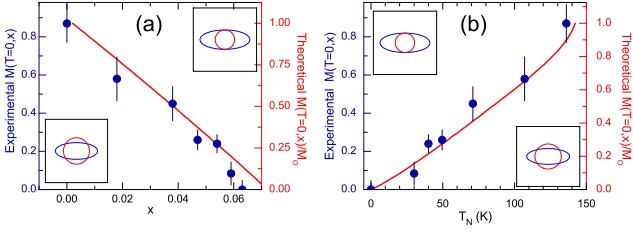} 
\par\end{centering}

\caption{Extrapolated zero temperature ordered moment $M\left(T=0,x\right)$
as function of doping $x$ (panel \textbf{a}) and as function of $T_{\mathrm{N}}$
(panel \textbf{b}) for Ba(Fe$_{1-x}$Co$_{x}$)$_{2}$As$_{2}$. Points
correspond to experimental data, whereas the solid line is the result
of the calculation described in the text. In the insets, the red circle
(blue ellipse) denotes the hole (electron) Fermi pocket.}

\end{figure}

\emph{Neutron diffraction experiments:} the neutron diffraction measurements
were performed on the HB1A diffractometer at the High Flux Isotope
Reactor at Oak Ridge National Laboratory on a series of Ba(Fe$_{1-x}$Co$_{x}$)$_{2}$As$_{2}$
single crystals using the same spectrometer configuration and data
analysis methods described in \cite{Pratt09}. The magnetic integrated
intensities were determined from rocking scans through the magnetic
peak at $\mathbf{Q}_{\mathrm{tet}}=\left(\frac{1}{2},\frac{1}{2},3\right)$
as a function of temperature and put on an absolute basis using the
known mass of the samples and the magnetic diffraction from the parent
compound, BaFe$_{2}$As$_{2}$, measured under identical conditions. 

The magnetic moment at zero temperature in the absence of SC, $M\left(T=0,x\right)$,
shown in Fig. 1, was determined by extrapolating the measured order
parameter $M\left(T,x\right)$  above $T_{c}$ using a power law fit to the data. The ratios
of the integrated intensities of the $\left(\frac{1}{2},\frac{1}{2},1\right)$
and $\left(\frac{1}{2},\frac{1}{2},3\right)$ magnetic reflections
were monitored to ensure that there was no change in the moment direction
as a function of temperature and composition $x$. No additional reflections,
e.g. incommensurate magnetic satellites, were observed, in agreement
with other work \cite{Lester09}. In Figs. 2a and b, we present a
systematic study of the temperature and composition dependence of
the magnetic order by neutron diffraction. The key result is shown
in Fig. 2b where we see that the magnetic order parameter, for superconducting
samples, peaks at $T_{c}$ and decreases for $T<T_{c}$. Indeed, for
$x=0.059$, our data show that there is a reentrance of the paramagnetic
phase (magnetic long-range order is completely suppressed), in agreement
with the predictions of our theory (see below). The opening of the
superconducting gap removes states at the Fermi surface that otherwise
contribute to the ordered moment, leading to a reduction of the ordered
moment below $T_{c}$.

\emph{Magnetic order in the absence of superconductivity}: in Fig.
1, we demonstrate that our model of itinerant magnetism provides a
description of the magnetically ordered state that is consistent with
the neutron diffraction data. The calculation of $M\left(T=0,x\right)$
is done by setting $\Delta_{l}=0$, but with magnetic order caused
by an electron-electron interaction $I\simeq0.95$ eV, chosen to yield
$T_{\mathrm{N}}=140$ K at $x=0$. The other parameters used were
$\varepsilon_{1,0}=0.095$ eV, $\varepsilon_{2,0}=0.125$ eV, $m=1.32m_{\mathrm{electron}}$,
$m_{x}=2m$ and $m_{y}=0.3m$, which yield an evolution of the Fermi
surface with doping consistent with what is seen by angle resolved photoemission spectroscopy (ARPES)
\cite{Liu09}. The suppression of AFM upon doping arises from the
detuning of the two Fermi surface sheets; the hole sheet around the
Brillouin zone center shrinks and the electron sheet grows upon electron
doping (see insets of Fig. 1). The carrier density, $x$, is fixed
under the assumption that each $\mathrm{Co}$ adds one electron. This
analysis fixes all our parameters, except for the pairing interaction
$V$. The latter is chosen to yield $T_{\mathrm{c}}\simeq25\mathrm{K}$
for $x$-values where AFM vanishes, yielding $\left\vert V \right\vert\simeq0.46$ eV at
$x\simeq0.062$.

\emph{Competing order:} without AFM order, many properties of $s^{++}$
and $s^{+-}$ pairing states are quite similar. This changes dramatically
once AFM and SC compete for the same electrons. For $s^{+-}$-pairing
the excitation energies $E_{\mathbf{p}\pm}$ are fully gapped, whereas
for the $s^{++}$-state nodes occur once $\Delta_{\mathrm{AFM}}>\Delta_{1}\Delta_{2}$,
at momentum values $\mathbf{p}_{n}$ given by $E_{\mathbf{p}_{n}-}=0$
\cite{Parker092}. This is true even if the nonmagnetic SC state
is fully gapped. For the case where $\Delta_{1}=\Delta_{2}$, nodes
are located at $\varepsilon_{\mathbf{p}_{n}\mathbf{,}1}=\varepsilon_{\mathbf{p}_{n}\mathbf{+Q},2}$
(i.e. where Bragg scattering due to AFM is large, see Ref.\cite{Parker092}).
Note, however, that in general nodes are not guaranteed to emerge:
for example, in the nested case $\varepsilon_{\mathbf{p,}1}=-\varepsilon_{\mathbf{p}+\mathbf{Q,}2}$
with $I>V$, the $s^{++}$-state remains fully gapped. To judge whether
these AFM-induced nodes for $s^{++}$-pairing are relevant, one needs
to analyze the free energy of Eq.\ref{F}, which determines whether
or not the two phases are allowed to coexist.

\begin{figure}
\begin{centering}
\includegraphics[width=1\columnwidth]{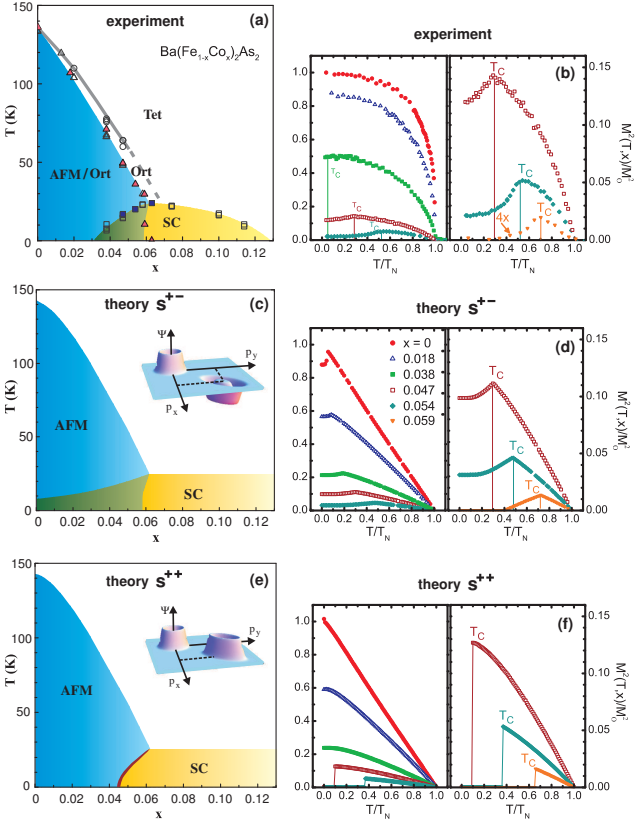} 
\par\end{centering}

\caption{\textbf{a,} The phase diagram of Ba(Fe$_{1-x}$Co$_{x}$)$_{2}$As$_{2}$
determined from neutron diffraction (solid symbols) as well as bulk
thermodynamic and transport measurements (open symbols, Ref.\cite{Ni09}).
\textbf{b,} The AFM order parameter squared measured via neutron diffraction
as function of temperature. The right-hand panel shows data in an expanded scale. $M_{0}=0.87\mu_{B}$ is the ordered moment
at $T=x=0$. \textbf{c,} and \textbf{d}, The phase diagram and theoretical
ordered moment, obtained for an unconventional $s^{+-}$ pairing state
(inset). Note the SC and AFM coexistence region, in green. Panels
\textbf{e} and \textbf{f} are analogous to \textbf{c} and \textbf{d,}
but for conventional $s^{++}$ pairing (see inset). Heterogeneous
coexistence of AFM and SC regions at the first order transition occurs
in panel (e) in a very narrow regime (dark red).}

\end{figure}

In Fig.2 we compare our theoretical results for the phase diagram
and the temperature dependence of $M^{2}$ with experiments. In Figs.
2c and d we show the calculated phase diagram and the behavior of
$M^{2}$ for the $s^{+-}$-state, using the parameters discussed above.
The phase coexistence and detailed temperature dependence of the ordered
moment agree well with experiment, including the narrow doping regime
with reentrance of the the AFM transition line. This is clearly different
for the $s^{++}$-state: in Fig.2e and f we show that the two phases
can not homogeneously coexist and are separated by a first order phase
transition. Thus, $s^{+-}$-pairing and magnetic order compete but
coexist microscopically, whereas both phases are mutually exclusive
in case of $s^{++}$-pairing.

This conclusion is robust and independent of specific details of the
model, as follows from a Landau expansion of Eq.\ref{F} with respect
to the order parameters. This expansion is performed near the multicritical
point $T_{\mathrm{N}}=T_{\mathrm{c}}$, where both phase lines meet
and where the decision about coexistence versus mutual exclusion takes
place. Figs. 3a, b and c show the Feynman diagrams that are responsible
for the order parameters coupling coefficients. The key diagram that
is responsible for the different behavior of $s^{++}$ and $s^{+-}$
pairing in Fig.2 is shown in Fig. 3c, corresponding to a term $\mathbf{M}^{2}\left(\Psi_{1}^{\ast}\Psi_{2}+\Psi_{2}^{\ast}\Psi_{1}\right)$
in the energy, which is sensitive to the relative phase, $\theta$,
between $\Psi_{1}$ and $\Psi_{2}$. Two partners of a Cooper pair
in one band are coherently scattered into the other band where they
recombine. AFM is essential, as it supplies the momentum transfer
$\mathbf{Q}$ needed for the scattering process. While the process
contributes to the total energy for either pairing state, the phase
of the pair wave-function determines the sign of this contribution,
causing the sensitivity of the phase diagram with respect to the internal
structure of the Cooper pair wave function.

To illustrate the physical origin (and generality) of our results,
we first discuss a simple limit that allows for analytic treatment.
Since the low energy electronic structure of the iron pnictides is
nearly particle-hole symmetric, we consider $T_{\mathrm{N}}=T_{\mathrm{c}}$
and assume particle-hole symmetry, i.e. $\varepsilon_{\mathbf{p}}\equiv\varepsilon_{\mathbf{p,}1}=-\varepsilon_{\mathbf{p}+\mathbf{Q,}2}$,
implying $\Psi\equiv\Psi_{1}=\mathrm{e}^{i\theta}\Psi_{2}$ and $I=\left\vert V \right\vert$.
In this limit, the Landau expansion of Eq.\ref{F} (relative to the
nonmagnetic normal state) yields:\begin{equation}
F=\frac{a}{2}\left(\left\vert \Psi\right\vert ^{2}+\mathbf{M}^{2}\right)+\frac{u}{4}\left(\left\vert \Psi\right\vert ^{2}+\mathbf{M}^{2}\right)^{2}+g\left(\theta\right)\left\vert \Psi\right\vert ^{2}\mathbf{M}^{2},\label{Landaus}\end{equation}
 which is highly symmetric in the two order parameters. $F$ depends
on the phase $\theta$ through $g\left(\theta\right)=\frac{u}{2}\left(1+\cos\theta\right)$,
and on the two coefficients $a=2I-\frac{I^{2}}{N}\sum\limits _{\mathbf{p}}\tanh\left(\frac{\varepsilon_{\mathbf{p}}}{2T}\right)/\varepsilon_{\mathbf{p}}$
and $u=\frac{I^{4}}{4NT}\sum\limits _{\mathbf{p}}\mathrm{sech}^{2}\left(\frac{\varepsilon_{\mathbf{p}}}{2T}\right)\left[T\:\sinh\left(\frac{\varepsilon_{\mathbf{p}}}{T}\right)-\varepsilon_{\mathbf{p}}\right] /\varepsilon_{\mathbf{p}}^{3}>0$.
Therefore, the following results are completely independent of further
details of the band structure dispersion, allowing us to draw general
conclusions about the phase diagram for different microscopic pairing
states.

If $g>0$, the two ordered states are separated by a first order transition,
while homogeneous phase coexistence and second order transitions only
occur if $g<0$. The $s^{++}$ state, with $g_{++}\equiv g\left(\theta=0\right)=u$,
is deep in the first order transition regime. Interestingly,
for particle-hole symmetry, the $s^{+-}$ state, with $g_{+-}\equiv g\left(\theta=\pi\right)=0$,
is at the border between regimes of coexistence and exclusion. The
special symmetry in Eq.\ref{Landaus} at $g=0$ is directly related
to the emergent SO(6) symmetry that was found in electronic
theories for $s^{+-}$ pairing\cite{Chubukov08} , as discussed in
Ref.\cite{Podolski09}, suggesting that this result holds beyond
weak-coupling. This demonstrates that, for the pnictides, $s^{+-}$-SC
is compatible with AFM, while magnetic order enforces new gap-nodes
or strongly reduces the gap for $s^{++}$-pairing\cite{Parker092},
impeding phase coexistence. It is also interesting to note the connection
of expression (\ref{Landaus}) with the SO(5) model proposed for
the cuprates \cite{Zhang97}.

Now, moving away from the special case of particle-hole symmetry,
we note that the inclusion of an infinitesimal chemical potential $\mu$
or a small ellipticity brings $g_{+-}$ to small but positive values.
However, when both the ellipticity and $\mu$ are finite, $g_{+-}$
can be negative. Using the parameters that lead to good agreement
with the experimental results in Figs. 1 and 2, we find virtually
the same result as before for $s^{++}$, $\frac{g_{++}}{u}\approx1$,
while for $s^{+-}$ the coefficient assumes a negative value, $\frac{g_{+-}}{u}\approx-0.26$,
allowing for the phase coexistence presented in Fig. 1.

We note that, for a $d$-wave state at particle-hole symmetry, we
obtain $\frac{g}{u}=\sqrt{\frac{2}{3}}-\frac{1}{2}$, i.e. it is less
compatible with AFM than the $s^{+-}$ state. A similar result is obtained 
for the case of a nodal $s^{+-}$ state. Even though our neutron
diffraction measurements did not detect any incommensurability, we
checked that our main results still hold even for a small incommensurability.
Furthermore, an extension of our model including the lattice degrees
of freedom satisfactory describes the behavior of the orthorhombic
state below $T_{c}$, as we show in \cite{Nandi10}.

\begin{figure}
\begin{centering}
\includegraphics[width=1\columnwidth]{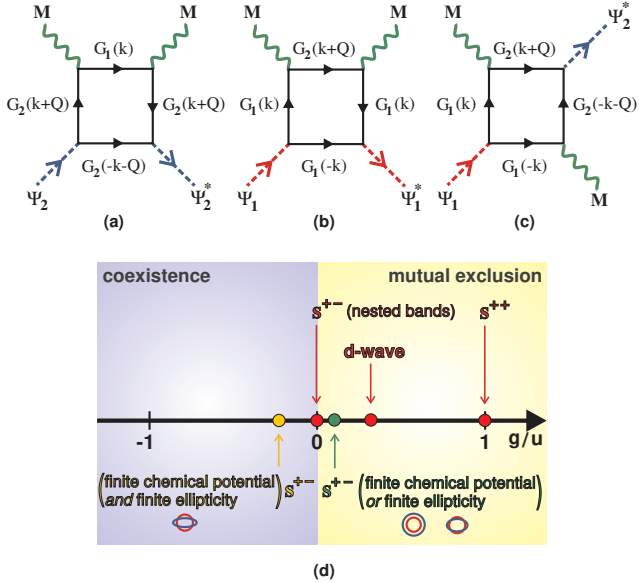} 
\par\end{centering}

\caption{\textbf{a, b, }and\textbf{ c,} Feynman diagrams responsible for the
coupling coefficients between the SC and AFM order parameters. $G_{i}\left(k\right)$
denotes the non-interacting single particle Green's function. Note
that diagram \textbf{c} is sensitive to the relative phase between
$\Psi_{1}$ and $\Psi_{2}$. \textbf{d,} Summary of the results for
the coupling coefficient $g$, considering different band dispersions.}

\end{figure}

The fact that the $s^{+-}$ state is on the verge of coexistence and
mutual exclusion with magnetism implies that the observation of different
phase diagrams, with and without coexistence\cite{Luetkens,Park09,Goko09,Ni09,Chu09,Ning09},
does not imply different pairing states. There are situations where
the $s^{+-}$ state coexists with AFM, as in Ba$\left(\text{Fe}_{1-x}\text{Co}_{x}\right)_{2}$As$_{2}$,
and others where it does not - as, for example, in the case of two
detuned circular bands \cite{Vorontsov09} (see Fig. 3d). Thus, the
presence of a first order transition does not imply $s^{++}$-pairing,
whereas the inverse is true: observing phase coexistence in the iron
arsenides disallows $s^{++}$ SC. This makes Ba$\left(\text{Fe}_{1-x}\text{Co}_{x}\right)_{2}$As$_{2}$
a crucially important member of the pnictide family.

We thank Alfred Kracher for performing the WDS measurements and Sergey
Bud'ko for assistance with the thermodynamic and transport measurements.
This work was supported by the U.S. DOE, Office of BES, DMSE. Ames
Laboratory is operated for the U.S. DOE by Iowa State University under
Contract No. DE-AC02-07CH11358.

\end{document}